\documentstyle[11pt,epsf]{article}
\newcommand {\dfn} {\stackrel{\Delta} {=}}

\newcommand {\bu} {\mbox{\boldmath $u$}}

\newcommand {\bx} {\mbox{\boldmath $x$}}
\newcommand {\hx} {\hat{x}}
\newcommand {\hH} {\hat{H}}
\newcommand {\hbx} {\hat{\mbox{\boldmath $x$}}}
\newcommand {\by} {\mbox{\boldmath $y$}}

\newcommand {\bE} {\mbox{\boldmath $E$}}

\newcommand{\calA}{{\cal A}}

\newcommand{\calE}{{\cal E}}

\newcommand{\calI}{{\cal I}}

\newcommand{\calK}{{\cal K}}

\newcommand{\calS}{{\cal S}}
\newcommand{\calT}{{\cal T}}

\newcommand{\calX}{{\cal X}}
\newcommand{\calY}{{\cal Y}}

\begin{document}
\thispagestyle{empty}
\title{Refinements and Extensions of Ziv's Model of Perfect Secrecy for
Individual Sequences}

\author{Neri Merhav}
\date{}
\maketitle

\begin{center}
The Andrew \& Erna Viterbi Faculty of Electrical and Computer Engineering\\
Technion - Israel Institute of Technology \\
Technion City, Haifa 3200003, ISRAEL \\
E--mail: {\tt merhav@technion.ac.il}\\
\end{center}
\vspace{1.5\baselineskip}
\setlength{\baselineskip}{1.5\baselineskip}

\begin{abstract}
We refine and extend Ziv's model and results regarding perfectly secure encryption of
individual sequences. According to this model, the encrypter and the
legitimate decrypter share in common a secret key, not shared with the unauthorized
eavesdropper, who is aware of the encryption scheme and has some prior
knowledge concerning the individual plaintext source sequence. 
This prior knowledge, combined with the cryptogram, is harnessed by
eavesdropper which implements a
finite-state machine as a mechanism for accepting or rejecting attempted guesses of
the source plaintext. The encryption is considered perfectly secure if the
cryptogram does not provide any new information to the eavesdropper that
may enhance its knowledge concerning the plaintext beyond his prior knowledge.
Ziv has shown that the key rate needed for perfect secrecy is essentially
lower bounded by the finite-state compressibility of the plaintext sequence, a bound
which is clearly asymptotically attained by Lempel-Ziv compression followed by
one-time pad encryption.
In this work, we consider some more general classes of finite-state
eavesdroppers and derive the respective lower bounds on the key rates needed
for perfect secrecy. These bounds are tighter and more refined than Ziv's bound and
they are attained by encryption schemes that are based on different universal
lossless compression schemes. We also extend our findings to the case where
side information is available to the eavesdropper and the legitimate
decrypter, but may or may not be available to the encrypter as well.
\end{abstract}

\section{Introduction}

Theoretical frameworks focusing on individual sequences and finite-state
encoders and decoders have undergone extensive exploration, diverging from the
conventional probabilistic paradigm used in modeling sources and channels.
This divergence has been particularly evident across various
information-theoretic domains, such as data compression
\cite{KY93}, \cite{YK96}, \cite{MZ06}, \cite{RM11}, \cite{WMF94}, \cite{WM02},
\cite{Ziv78}, \cite{Ziv80}, \cite{Ziv84}, \cite{ZL78}, 
source/channel simulation
\cite{MMSW10}, \cite{Seroussi06}, 
classification \cite{Ziv88}, \cite{ZM93}, \cite{Merhav00}, prediction 
\cite{FMG92}, \cite{HKW98}, \cite{MF93}
\cite{WM01}, \cite{WMSB01},
\cite{ZM07}, 
denoising \cite{WOSVW05}, and
even channel coding \cite{LF11}, \cite{LF13}, \cite{SF05}. 
These references merely scratch the surface
of a vast body of literature.
Conversely, the realm of information-theoretic security, from Shannon's
seminal contribution \cite{Shannon48} to more contemporary research 
\cite{Hellman77},
\cite{Lempel79}, \cite{LPS09}, \cite{Massey88}, \cite{Yamamoto91},
remains almost totally entrenched in the probabilistic framework. While these works
represent a mere fraction of the extensive literature, they exemplify the
nearly exclusive reliance on probabilistic models within this field.

To the best of the author's knowledge, there are only two exceptions to this
prevailing paradigm, documented in an unpublished memorandum by Ziv
\cite{Ziv78u} and a
subsequent work \cite{me13}. Ziv's memorandum presents a unique approach wherein the
plaintext source, to be encrypted using a secret key, is treated as an
individual sequence. The encrypter is modeled as a general block encoder,
while the eavesdropper employs a finite-state machine (FSM) as a message
discriminator. That memorandum postulates that the eavesdropper possesses
certain prior knowledge about the plaintext, expressed as a set of
``acceptable messages'', hereafter referred to as the {\em acceptance set}.
In other words, before observing the cryptogram, the eavesdropper
uncertainty about the plaintext sequence is that it could be any member is in
this set of acceptable messages. 

This assumption about prior knowledge available to the eavesdropper is fairly
realistic in real life. 
Consider, for example, the case where 
the plaintext alphabet is the latin alphabet, but
the eavesdropper furthermore knows that the plaintext
must be a piece of text in the
Italian language. In this case, her prior knowledge, first and foremost, allows her to
reject every candidate string of symbols that includes 
the letters `j', `k', `w', `x' and `y', 
which are not used in Italian. Another example, which is common to English and some
other languages, is that the letter `q' must
be followed by `u'. In the same spirit, some additional rules of grammar can
be invoked, like limitations on 
the number of successive consonant (or vowel) letters
in a word, a limitation on the length of a word, and so on.

Now, according to Ziv's approach, perfectly secure encryption amounts
to a situation where the presence of the cryptogram does not reduce the
uncertainty associated with the acceptance set. In other words, having intercepted the cryptogram, the
eavesdropper learns nothing about the
plaintext that she did not not know before.
The size of the acceptance set can be thought of as a quantifier of the level of uncertainty: a larger
set implies greater uncertainty. The aforementioned FSM is used to
discriminate between acceptable and unacceptable strings of plaintext symbols
that can be obtained by examining various key bit sequences. Accordingly, perfect security 
amounts to maintaining the size of the acceptance set unchanged, and
consequently, the uncertainty level, in the presence of the cryptogram.
The principal finding in Ziv's work is that the asymptotic key rate required
for perfectly secure encryption, according to this definition, cannot be lower
(up to asymptotically vanishing terms) than the Lempel-Ziv (LZ) complexity of
the plaintext source \cite{ZL78}. Clearly, this lower bound is asymptotically
achieved by employing one-time pad encryption (that is, bit-by-bit XOR with
key bits) of the bit-stream obtained from
LZ data compression of the plaintext source, mirroring Shannon's classical
probabilistic result which asserts that the minimum required key rate equals
the entropy rate of the source.

In the subsequent work \cite{me13}, the concept of perfect secrecy for individual
sequences was approached differently. Instead of a finite-state eavesdropper
with predefined knowledge, it is assumed that the encrypter can be realized by
an FSM which is sequentially fed by the plaintext source and random key bits. A notion
of ``finite-state encryptability'' is introduced (in the spirit of the analogous finite-state
compressibility of \cite{ZL78}), which designates the minimum key rate which
must be consumed by any finite-state encrypter, such that probability law of
the cryptogram would be independent of the plaintext input, and hence be perfectly
secure. Among the main results of \cite{me13}, it is asserted and proved that
the finite-state encryptability of an individual sequence is essentially
bounded from below by its finite-state compressibility, a bound which is once again
attained asymptotically by LZ compression followed by one-time pad encryption.

In this work, we revisit Ziv's approach to perfect secrecy for individual
sequences \cite{Ziv78u}. After presenting his paradigm in detail, we
proceed to refine and generalize his findings in certain aspects. First, we consider
several more general classes of finite-state discriminators that can be employed 
by the eavesdropper. These will lead to tight lower bounds on the minimum key
rate to be consumed by the encrypter, which will be matched by encryption
schemes that are based some other universal data compression schemes. The
resulting gaps
between the lower bounds and the corresponding upper bounds (i.e., the
redundancy rates) would converge faster. Among these more general classes of
finite-state machines, we will consider finite-state machines that are
equipped with counters, as well as periodically time-varying finite-state
machines with counters. Another direction of generalizing Ziv's findings is
the incorporation of side information (SI) that is available both at the
eavesdropper and the legitimate decrypter, but may or may not be available at
the encrypter as well.

The outline of this article is as follows.
In Section \ref{fnb}, we formulate the model setting, establish the notation, and
provide a more detailed background on Ziv's model and results in \cite{Ziv78u}.
In Section \ref{main}, which is the main section of this article, we present
the refinements and extensions to other types of FSMs, including FSMs with
counters (Subsection \ref{fsm+c}), shift-register FSMs with counters
(Subsection \ref{srm+c}), and periodically time-varying
FSMs with counters (Subsection \ref{ptv+c}). Finally, in Section \ref{si}, we further extend some of
our findings to the case where SI is available at both the legitimate
decrypter and the eavesdropper, but not necessarily at the encrypter.

\section{Formulation, Notation and Background}
\label{fnb}

Consider the following version of Shannon's cipher system model, adapted to
the entryption of individual sequences, as proposed by Ziv \cite{Ziv78u}.
An individual (deterministic)
plaintext sequence, $\bx=(x_0,\ldots,x_{n-1})$ ($n$ - positive integer), is encrypted using a random key $K$,
whose entropy is $H(K)$, to generate a cryptogram, $W=T(\bx,K)$, where the
mapping $T(\cdot,K)$ is invertible given $K$, namely, $\bx$ can be reconstructed by the
legitimate decoder, who has access to $K$, by applying the inverse
function, $\bx=T^{-1}(W,K)$. The plaintext symbols, $x_i$,
$i=0,1,2,\ldots,n-1$,
take on values in a finite alphabet, $\calX$, of size $\alpha$.
Thus, $\bx$ is a member of $\calX^n$, the $n$-th Cartesian power of $\calX$,
whose cardinality is $\alpha^n$. Without essential loss of generality, we
assume that $K$ is a uniformly distributed random variable taking on values in
a set $\calK$ whose cardinality is $2^{H(K)}$. 
Sometimes, it may be convenient to consider $\calK$ to be the set
$\{0,1,\ldots, 2^{H(K)}-1\}$.
A specific realization of the
key, $K$, will be denoted by $k$. 

An eavesdropper, who knows the mapping $T$, but not the realization of the
key, $K$, is in the quest of learning as much as
possible about $\bx$ upon observing $W$. It is assumed that the eavesdropper
also has some prior knowledge about $\bx$, even before observing $W$. In particular, the
eavesdropper knows that the plaintext source string $\bx$ must be a member of
a certain subset of $\calX^n$, denoted $\calA_n$, which is referred to as the
{\em acceptance set}. 

Ziv models the eavesdropper by a
cascade of a guessing decrypter and a finite-state message discriminator, which
work together as follows. At each step, the eavesdropper examines a certain
key, $k\in\calK$, by generating an estimated plaintext,
$\hat{\bx}=T^{-1}(W,k)$, and then feeding $\hat{\bx}$ into the message
discriminator to examine whether or not $\hat{\bx}\in\calA_n$. If the answer is
affirmative, $\hat{\bx}$ is accepted as a candidate, otherwise, it is
rejected. Upon completing this step, the eavesdropper moves on to the next
key, $k+1$, and repeats the same process, etc. The message discriminator is modeled as a finite-state
machine, which implements the following recursive equations for
$i=0,1,2,\ldots,n-1$:
\begin{eqnarray}
u_i&=&f(z_i,\hat{x}_i)\\
z_{i+1}&=&g(z_i,\hat{x}_i),
\end{eqnarray}
where $z_0,z_1,z_2,\ldots,z_{n-1}$ is a sequence
of states, $z_i\in\calS$, $i=0,1,2,\ldots,n-1$, $\calS$ being a set of $s$ states
(and with the initial state, $z_0$, as a fixed member of $\calS$),
$u_0,u_1,\ldots,u_{n-1}$ is a binary output sequence,
$f:\calS\times\calX\to\{0,1\}$ is the output function, and
$g:\calS\times\calX\to\calS$ is the next-state function. If
$u_0,u_1,u_2,\ldots,u_{n-1}$ is the all-zero sequence, $\hbx$ is accepted,
namely, $\hbx\in\calA_n$, otherwise,
as soon as $u_i=1$ for some $0\le i\le n-1$, $\hat{\bx}$ is rejected. In other
words, $\calA_n$ is defined to be the set of all $\{\hbx\}$ for which the response of the
finite-state discriminator is the all-zero sequence, $\bu=(0,0,\ldots,0)$.\\

\noindent
{\bf Example 1.}
Let $\calX=\{0,1\}$ and suppose that membership of $\bx$ in $\calA_n$ forbids the appearance of more
than two successive zeroes. Then, a simple
discriminator can detect the violence of this rule using the 
finite-state machine defined by $\calS=\{0,1,2\}$ and
\begin{eqnarray}
g(z,\hat{x})&=&\left\{\begin{array}{ll}
(z+1)~\mbox{mod}~3 & \hat{x}=0\\
0 & \hat{x}=1\end{array}\right.\\
f(z,\hat{x})&=&\left\{\begin{array}{ll}
1 & z=2~\mbox{and}~\hat{x}=0\\
0 & \mbox{otherwise}\end{array}\right.
\end{eqnarray}
More generally, consider the set of binary sequences that comply with
the so called $(d,k)$-constraints, well known from the literature of magnetic
recording (see, e.g., \cite{MRS98} and references therein), namely, binary sequences, where the runs of successive zeroes
must be of length at least $d$ and at most $k$, where $d$ and $k$ (not to be
confused with the notation of the encryption key) are positive
integers with $d\le k$. We shall return to this example later. 
$\Box$\\

Ziv defines perfect secrecy for individual sequences as a situation where even
upon observing $W$,
the eavesdropper's uncertainty about $\bx$ is not reduced. In the mathematical
language, let us denote
\begin{equation}
T^{-1}(W)\dfn\{T^{-1}(W,k),~k\in\calK\},
\end{equation}
and
\begin{equation}
\calA_n(W)\dfn\calA_n\bigcap T^{-1}(W),
\end{equation}
then, perfect secrecy is defined as a situation where
\begin{equation}
A_n(W)=A_n,
\end{equation}
or equivalently,
\begin{equation}
\label{setinequality}
\calA_n\subseteq T^{-1}(W).
\end{equation}
To demonstrate these concepts, consider the following example.\\

\noindent
{\bf Example 2.}
Let $n=4$, $\calX=\{0,1\}$, $\bx=(1 1 1 1)$, $2^{H(K)}=8$, $k=(1 1 1 1)$,
and then for a one-time pad encrypter,
\begin{equation}
W=T(\bx,k)=\bx\oplus k=(1 1 1 1)\oplus (1 1 1 1)=(0 0 0 0),
\end{equation}
where $\oplus$ denotes bit-wise XOR (modulo 2 addition). Let the set $\calK$ of all 8
possible key strings be given by
\begin{equation}
\calK=\left\{\begin{array}{cc}
1 1 1 1\\
1 0 0 0\\
1 1 0 0\\
1 0 0 1\\
0 0 0 0\\
0 1 1 1\\
0 0 1 1\\
0 1 1 0\end{array}\right\}.
\end{equation}
Obviously, the decryption is given by $T^{-1}(W,k)=W\oplus k$.
Since $W=(0 0 0 0)$, then $T^{-1}(W)=\calK$. Following Example 1, suppose that $\calA_4$ is the set of
all binary vectors of length $n=4$, which do not contain runs of more than two
zeroes. There are only 3 binary vectors of length 4 that contain a succession
of more than 2 (i.e., 3 or 4) zeroes, namely, $(0 0 0 0)$, $(1 0 0 0)$, and
$(0 0 0 1)$. Thus,
$|\calA_4|=2^4-3=13$. On the other hand,
\begin{equation}
|\calA_4(W)|=|\calA_4\bigcap T^{-1}(W)|\le|T^{-1}(W)|=|\calK|=8 < 13,
\end{equation}
which means
that this encryption system is not perfectly secure. 
The reason is that the key space, $\calK$, is not large enough.
$\Box$\\

Clearly, the best one can do in the quest of minimizing $H(K)$, without compromising perfect
secrecy, is to design the encrypter in
such a way that
\begin{equation}
T^{-1}(W)=\calA_n,
\end{equation}
for every cryptogram $W$ that can possibly be obtained from some combination
of $\bx$ and $k$.
Conceptually, this can be obtained by mapping $\calA_n$ to the set of all binary
sequences of length $H(K)$ by means of a fixed-rate data compression scheme
and applying one-time pad encryption to the compressed sequence.
Here, and throughout the sequel, we neglect integer length constraints
associated with large numbers, and so, $H(K)$ is assumed integer without
essential loss of generality and optimality.\\

\noindent
{\bf Remark 1.} For readers familiar with the concepts and the terminology of
coding for systems with $(d,k)$
constraints (and other readers may skip this remark without loss of
continuity), it is insightful to
revisit the second part of Example 1: 
If $\calA_n$ is the set of binary $n$-sequences that
satisfy a certain $(d,k)$ constraint, then optimal encryption for $\calA_n$
pertains to compression using the inverse mapping of a channel encoder for the same $(d,k)$
constraint, namely, the
corresponding channel decoder, which is followed by
one-time pad encryption. The minimum key rate needed is then equal to the
capacity of the constrained system, which can be calculated either
algebraically, as the logarithm of the Perron-Frobenius eigenvalue of the
state adjacency matrix of the state transition diagram, or probabilistcally, as the maximum entropy
among all stationary Markov chains that are supported by the corresponding state transition
graph \cite{MRS98}. $\Box$\\

Ziv's main result in \cite{Ziv78u} is that for a finite-state discriminator,
if $\bx\in\calA_n$, the cardinality of
$\calA_n$ cannot be exponentially smaller than $2^{LZ(\bx)}$ (see Appendix for
the proof), 
where $LZ(\bx)$ is the length (in bits) of the compressed version of $\bx$
using the 1978 version of the Lempel-Ziv algorithm (the LZ78 algorithm) \cite{ZL78},
and so, the
key rate needed in order to completely encrypt all members of $\calA_n$ is lower
bounded by
\begin{eqnarray}
\label{R>LZ}
R&\dfn&\frac{H(K)}{n}\nonumber\\
&=&\frac{\log|T^{-1}(W)|}{n}\nonumber\\
&\ge&\frac{\log|\calA_n|}{n}\nonumber\\
&\ge&\frac{LZ(\bx)}{n}-\epsilon_n
\end{eqnarray}
where $\epsilon_n$ is a positive sequence tending to zero as $n\to\infty$ at
the rate of $\frac{\log(\log n)}{\log n}$, and
where here and throughout the sequel, the notation $|\calE|$ for a finite set
$\calE$, designates the cardinality of $\calE$.
The first inequality of (\ref{R>LZ}) follows from (\ref{setinequality}). 
Obviously, this bound is essentially attained by
LZ78 compression of $\bx$, followed by one-time pad encryption using $LZ(\bx)$
key bits. As can be seen, the gap, $\epsilon_n$, between the upper bound and the lower bound
on $R$ is $O\left(\frac{\log(\log n)}{\log n}\right)$, which tends
to zero rather slowly.\\

\noindent
{\bf Remark 2.}
For an infinite sequence $\bx=x_0,x_1,x_2,\ldots$, asymptotic results are obtained
in \cite{Ziv78u} by a two-stage limit: First, consider a finite sequence of total length $m\cdot
n$, which is divided into $m$ non-overlapping $n$-blocks, where the above
described mechanism is applied to each $n$-block separately. The asymptotic
minimum key rate is obtained by a double limit superior, which is taken first,
for $m\to\infty$ for a given $n$, and then for $n\to\infty$. In this work, we
will have in mind a similar double limit, but we shall not mention it
explicitly at every relevant occasion. Instead, we will focus on the behavior
of a single $n$-block. 

\section{More General Finite-State Discriminators}
\label{main}

This section, which is the main section in this article, is devoted to describe several more
general classes of finite-state discriminators along with derivations of their respective
more refined bounds.

\subsection{FSMs with Counters}
\label{fsm+c}

While Ziv's model for a finite-state discriminator is adequate for rejecting
sequences with certain forbidden patterns (like a succession of more than
two zeroes in the above examples), it is not sufficient to handle
situations like the following one. Suppose that encrypter applies a universal
lossless compression algorithm for memoryless sources, followed by one-time
pad encryption. Suppose also that the universal compression scheme is a
two-part code, where the first part encodes the index of the type class of
$\bx$, using a number of bits that is proportional to $\log n$, 
and the second part represents the index of $\bx$ within the type
class, assuming that the encoder and the decoder have agreed
on some ordering ahead of time. In this case, the length of the cryptogram (in bits), which is equal to the
length of the compressed data, is about
\begin{equation}
L=H(K)\approx n\hat{H}(\bx)+\frac{\alpha-1}{2}\cdot\log n, 
\end{equation}
where $\hat{H}(\bx)$ is the empirical entropy of
$\bx$ (see, e.g., \cite{WMF94} and references therein). 
The eavesdropper, being aware of the encryption scheme, observes the
length $L$ of the cryptogram, and immediately concludes that $\bx$ must be
a sequence whose empirical entropy is 
\begin{equation}
H_0=\frac{1}{n}\cdot\left(L-\frac{\alpha-1}{2}\cdot\log n\right).
\end{equation}
In other words, in this case,
\begin{equation}
\calA_n=\left\{\hbx:~\hat{H}(\hbx)=H_0\right\}.
\end{equation}
Therefore, every sequence whose empirical distribution pertains to empirical
entropy different from $H_0$ should be rejected. To this end, our discriminator
should be able to gather empirical statistics, namely, to count occurrences of
symbols (or more generally, count combinations of symbols and states) and not just to detect a forbidden
pattern that might have occurred just once in $\bx$. 

This motivates us to broaden the class of finite-state discriminators to be
considered, in the following fashion. We consider discriminators that consist of a next-state
function,
\begin{equation}
z_{i+1}=g(z_i,\hat{x}_i),
\end{equation}
as before, but instead of the binary output function, $f$, of \cite{Ziv78u},
these discriminators are equipped with
a set of $\alpha\cdot s$ counters that count the number of joint
occurrences of all $(x,z)\in\calX\times\calS$ for $i=0,1,2,\ldots,n-1$, i.e.,
\begin{equation}
n(x,z)=\sum_{i=1}^n\calI\{\hx_i=x,~z_i=z\},~~~~~x\in\calX,~z\in\calS,
\end{equation}
where $\calI\{A\}$, for a generic event $A$, denotes its indicator function,
namely, $\calI\{A\}=1$ if $A$
is true and $I\{A\}=0$ if not. 
A sequence $\hbx$ is accepted (resp.\ rejected) if the matrix of counts,
$\{n(x,z),~x\in\calX,~z\in\calS\}$,
satisfies (resp.\ violates) a certain condition. In the example of the previous paragraph, a
sequence is accepted if
\begin{equation}
\hat{H}(\hbx)=\sum_{x\in\calX}\frac{n(x)}{n}\log\left(\frac{n}{n(x)}\right)=
H_0, 
\end{equation}
where $n(x)=\sum_{z\in\calS}n(x,z)$. Ziv's discriminator model is clearly a special
case of this model: let $(x_\star,z_\star)$ be any combination of input and state such that
$f(z_\star,x_\star)=1$ in Ziv's model (that is, a ``forbidden'' combination). 
Then, in terms of the proposed extended model, a sequence is
rejected whenever $n(x_\star,z_\star)\ge 1$.

Clearly, since $\bx\in\calA_n$ and since membership in $\calA_n$ depends solely on
the matrix of counts, $\{n(x,z),~x\in\calX,~z\in\calS\}$, it follows that all
$\{\hbx\}$ that share the same counts as these of $\bx$ must also be members
of $\calA_n$.
The set of all $\{\hbx\}$ of length $n$ with the same counts,
$\{n(x,z),~x\in\calX,~z\in\calS\}$, as these of $\bx$, 
is called the
{\em finite-state type class} w.r.t.\ the FSM $g$ (see also \cite{WMF94}), and
it is denoted by $\calT_g(\bx)$. Since $\calA_n\supseteq\calT_g(\bx)$,
\begin{equation}
\label{A>T}
|\calA_n|\ge |\calT_g(\bx)|.
\end{equation}
It is proved in
\cite{WMF94} (Lemma 3 therein), that if $n(x,z)\ge n(z)\delta(n)$ for every
$(x,z)\in\calX\times\calS$, where $\delta(n)>0$ may
even be a vanishing sequence, then
\begin{equation}
\label{T>H}
|\calT_g(\bx)|\ge \exp_2\left\{n\hat{H}(X|Z)-\frac{s(\alpha-1)}{2}\cdot\log
(2\pi n)\right\},
\end{equation}
where 
\begin{equation}
\hat{H}(X|Z)=-\sum_{x\in\calX}\sum_{z\in\calS}\frac{n(x,z)}{n}\log\frac{n(x,z)}{n(z)},
\end{equation}
with $n(z)=\sum_{x\in\calX}n(x,z)$ and with the conventions that $0\log
0\dfn 0$ and $0/0\dfn 0$. It therefore follows that the key rate needed for
perfect secrecy is lower bounded by
\begin{eqnarray}
R&=&\frac{H(K)}{n}\nonumber\\
&\ge&\frac{\log|\calA_n|}{n}\nonumber\\
&\ge&\frac{\log|\calT_g(\bx)|}{n}\nonumber\\
&\ge&\hat{H}(X|Z)-\frac{s(\alpha-1)}{2}\cdot\frac{\log(2\pi n)}{n}.
\end{eqnarray}
This lower bound can be asymptotically attained by an encrypter that applies
universal loss compression for finite-state sources with the next state
function $g$, followed by one-time pad encryption. This universal
lossless compression scheme is based on a conceptually simple extension of the
above-mentioned universal scheme for the class of memoryless sources: one applies a two-part code where the first part
includes a code for the index of the type class w.r.t.\ $g$ (using a number of
bits that is proportional to $\log n$), and the second part encodes the index
of the location of $\hbx$ within $\calT_g(\bx)$ according to a predefined order
agreed between the encoder and the decoder (see \cite{WMF94} for more
details). More precisely, the compression ratio that can be achieved, which is
also the key rate consumed, is upper bounded by (see Section III of
\cite{WMF94}):
\begin{equation}
R\le\hat{H}(X|Z)+\frac{s(\alpha-1)}{2}\cdot\frac{\log
n}{n}+O\left(\frac{1}{n}\right),
\end{equation}
which tells us that the gap between the lower bound and the achievability is
proportional to $\frac{\log n}{n}$, which decays much faster than the
aforementioned $O\left(\frac{\log(\log n)}{\log n}\right)$ convergence rate of the gap in Ziv's
approach. Moreover, for most sequences in most type classes, $\{\calT_g(\bx)\}$, 
the coding rate (which is also the key rate in one-time-pad encryption) of
$\hat{H}(X|Z)+\frac{s(\alpha-1)}{2}\cdot\frac{\log
n}{n}$ is smaller than $LZ(\bx)/n$ since the former quantity is essentially
also a lower bound to the compression ratio of any lossless compression scheme
for most individual sequences in $\calT_g(\bx)$, for almost all such type
classes \cite[Theorem 1]{WMF94}. The converse inequality between
$\hat{H}(X|Z)$ and $LZ(\bx)/n$, which follows from Ziv's
inequality (see
\cite[Lemma 13.5.5]{CT06} and \cite{PWZ92}) holds up to an
$O\left(\frac{\log(\log n)}{\log n}\right)$ term again.\\

\noindent
{\bf Remark 3.}
To put Ziv's result in perspective, one may wish to envision a class of
discriminators defined by a given dictionary of $c$ distinct `words', which
are the $c$ phrases in Ziv's derivation \cite{Ziv78u} (see also Appendix).
A possible definition of such a discriminator is that it accepts only
$n$-sequences of plaintext that are formed by concatenating words from
the given dictionary, allowing repetitions. 
These are no longer finite-state discriminators. In this case, Ziv's
derivation is applicable under the minor modification that the various phrases
should be classified only in terms of their length, but without further partition according
to initial and final states ($z$ and $z'$ in the derivation of the appendix).
This is equivalent to assuming that $s=1$, and
accordingly, the term $\frac{2c\log s}{n}$ in the last equation of the
appendix should be omitted. Of course, the resulting bound can still
be matched by LZ78 compression followed by one time-pad encryption.

\subsection{Shift-Register Machines with Counters}
\label{srm+c}

Since the eavesdropper naturally does not cooperate with the encrypter,
the latter might not know the particular FSM, $g$, used by the former, and
therefore, it is instructive to derive key-rate bounds that are independent of
$g$. To this end, consider the following.
Given $\bx$ and a fixed positive integer $\ell$ ($\ell\ll n$), let us observe
the more refined joint empirical distribution,
\begin{equation}
\hat{P}_{X^{\ell+1}Z^{\ell+1}}(a_0,\ldots,a_\ell,s_0,\ldots,s_\ell)=\frac{1}{n}\sum_{i=0}^{n-1}
\calI\{x_{i}=a_0,\ldots,x_{i\oplus\ell}=a_\ell,
z_{i}=s_0,\ldots,z_{i\oplus\ell}=s_\ell\},
\end{equation}
as well as all partial marginalizations derived from this distribution, where
here, $\oplus$ denotes addition modulo $n$ so to create a periodic extension
of $\bx$ (hence redefining $z_0=g(z_{n-1},x_{n-1})$). Accordingly, the previously defined
empirical conditional entropy, $\hat{H}(X|Z)$ is now denoted
$\hat{H}(X_1|Z_1)$, which is also equal to 
$\hat{H}(X_2|Z_2)$, etc., due to the inherent shift-invariance property of the
empirical joint distribution extracted under the periodic extension of $\bx$.
Consider now the following chain of inequalities:
\begin{eqnarray}
& &\hat{H}(X_\ell|X_0,X_1,\ldots,X_{\ell-1})-
\hat{H}(X_\ell|Z_\ell)\nonumber\\
&\le&\frac{1}{\ell+1}\sum_{j=0}^{\ell}[\hat{H}(X_j|X_0,\ldots,X_{j-1})-\hat{H}(X_j|X_0,\ldots,X_{j-1},Z_0)]\\
&=&\frac{1}{\ell+1}[\hat{H}(X_0,\ldots,X_\ell)-\hat{H}(X_0,\ldots,X_\ell|Z_0)]\\
&=&\frac{\hat{I}(Z_0;X_0,\ldots,X_\ell)}{\ell+1}\\
&\le&\frac{\hat{H}(Z_0)}{\ell+1}\\
&\le&\frac{\log s}{\ell+1},
\end{eqnarray}
where $\hat{I}(\cdot;\cdot)$ denotes empirical mutual information and where
in the second line, the term corresponding to $j=0$ should be understood
to be $[\hat{H}(X_0)-\hat{H}(X_0|Z_0)]$.
The first inequality follows because given $(X_0,\ldots,X_{j-1},Z_0)$,
one can reconstruct $Z_1,Z_2,\ldots,Z_j$ by $j$ recursive applications of the next-state
function, $g$. Therefore, 
\begin{eqnarray}
\hat{H}(X_j|X_0,\ldots,X_{j-1},Z_0)&=&
\hat{H}(X_j|X_0,\ldots,X_{j-1},Z_0,Z_1,\ldots,Z_j)\nonumber\\
&\le&\hat{H}(X_j|Z_j)\\
&=&\hat{H}(X_\ell|Z_\ell).
\end{eqnarray}
Equivalently,
\begin{equation}
\hat{H}(X_\ell|Z_\ell)\ge
\hat{H}(X_\ell|X_0,\ldots,X_{\ell-1})-\frac{\log s}{\ell+1},
\end{equation}
and so, combining this with eqs.\ (\ref{A>T}) and (\ref{T>H}), we get
\begin{equation}
\log|\calA_n|\ge n\left[\hat{H}(X_\ell|X_0,\ldots,X_{\ell-1})-\frac{\log
s}{\ell+1}\right]-\frac{s(\alpha-1)}{2}\cdot\log(2\pi n).
\end{equation}
The advantage of this inequality is in its independence upon the
arbitrary next-state function $g$. In fact, we actually replaced the arbitrary
FSM, $g$, by a particular FSM -- the
shift-register FSM, whose state is
$z_i=(x_{i-\ell},x_{i-\ell+1},\ldots,x_{i-1})$ at the cost of a gap of $\frac{\log
s}{\ell+1}$, which can be kept arbitrarily small if the size of the
shift-register, $\ell$, is sufficiently large compared to the memory size,
$\log s$, of $g$.\\

\noindent
{\bf Remark 4.} The fact that $\hat{H}(X_\ell|X_0,\ldots,X_{\ell-1})$ cannot
be much larger than $\hat{H}(X_\ell|Z_\ell)$ for large enough $\ell$
actually suggests that whatever the state of any FSM $g$ can possibly
``remember'' from the past of $\bx$ is essentially captured by the recent
past, and not
by the remote past. While this is not surprising in the context of the
probabilistic setting, especially if the underlying random process is ergodic
and hence has a vanishing memory of the remote past, this finding is not quite
trivial when it comes to arbitrary individual sequences.
$\Box$\\

Returning to our derivations,
in view of the first three lines of (\ref{R>LZ}),
the key rate needed for perfect secrecy is lower bounded by
\begin{equation}
R\ge \frac{\log|\calA_n|}{n}\ge\hat{H}(X_\ell|X_0,\ldots,X_{\ell-1})-\frac{\log
s}{\ell+1}-\frac{s(\alpha-1)}{2n}\cdot\log(2\pi n),
\end{equation}
and since this holds for any $\ell$ in some fixed range $1\le \ell\le l$
(i.e., where $l$ is independent of $n$),
\begin{equation}
R\ge \max_{1\le\ell\le l}\left[\hat{H}(X_\ell|X_0,\ldots,X_{\ell-1})-\frac{\log
s}{\ell+1}\right]-\frac{s(\alpha-1)}{2n}\cdot\log(2\pi n),
\end{equation}
Note that if $\bx$ is an ``$\ell_0$-th order Markovian sequence'' in the
sense that $\hat{H}(X_\ell|X_0,\ldots,X_{\ell-1})$ is almost fixed for all
$\ell_0\le\ell\le l$ with $l\gg\log s$, then $\ell_0$ is the preferred choice for
$\ell$ as it essentially captures the best attainable key rate.

This lower bound can be asymptotically attained by universal lossless
compression for $\ell$-th order Markov types \cite{DLS81}, \cite{Natarajan85},
\cite[Section VII.A]{Csiszar98}, followed by one-time pad
encryption, where the achieved rate is
\begin{equation}
R\le \hat{H}(X_\ell|X_0,\ldots,X_{\ell-1})
+\frac{\alpha^{\ell-1}(\alpha-1)}{2}\cdot\frac{\log n}{n}+O\left(\frac{1}{n}\right).
\end{equation}
In this case, $\calA_n$ is the $\ell$-th order Markov type of $\bx$ and the
finite-state discriminator of the eavesdropper is a shift-register machine,
that checks whether the $\ell$-th order Markov type of each $\hbx$ has the
matching conditional empirical conditional entropy of order $\ell$.

In this result, there is a compatibility between the converse bound and the
achievability bound in the sense that both are given in terms of FSMs
with a fixed number of states that does not grow with $n$. 
Among all possible finite-state machines, we have actually single out the shift-register
machine universally, at the cost of a controllable asymptotic gap of $\frac{\log
s}{\ell+1}$, but otherwise, the bound is explicit and it is clear how to
approach it. If we
wish to keep this gap below a given $\epsilon>0$, then we select $\ell=\lceil
(\log s)/\epsilon\rceil-1$. In this
sense, it is another refinement of Ziv's result. 

\subsection{Periodically Time-Varying FSMs with Counters}
\label{ptv+c}

So far, we have considered time-invariance FSM, where the function $g$ remains
fixed over time. We now expand the scope to
consider the class of discriminators implementable by
periodically time-varying FSMs, defined as follows:
\begin{eqnarray}
z_{i+1}&=&g(z_i,\hat{x}_i, i~\mbox{mod}~l),
\end{eqnarray}
where $i=0,1,2,\ldots$ and $l$ is a positive integer that designates the period of the
time-varying finite-state machine. Conceptually, this is not really more
general than the ordinary, time-invariant FSM defined
earlier, because the state of the modulo-$\ell$ clock can be considered part of
the entire state, in other words, this is a time-invariant FSM
with $s\cdot l$ states, indexed by the ordered pair $(z_i,
i~\mbox{mod}~l)$.
The reason it makes sense to distinguish between the state $z_t$
and the state of the clock is because the clock does not store any information
regarding past input data. This is to say that in the context of 
time-varying finite-state machines, we distinguish between the amount of memory
of past input ($\log s$ bits) and the period $l$. Both parameters manifest
the richness of the class of machines, but in different manners. 
Indeed, the parameters $s$ and $l$ will play very different and separate roles in the
converse bound to be derived below.\\

\noindent
{\bf Remark 5.} Clearly, the earlier
considered time-invariant finite-state machine is obtained as the special case pertaining to
$l=1$, or to the case where $l$ is arbitrary, but the next-state functions, $g(\cdot,\cdot,0),
g(\cdot,\cdot,1),\ldots,g(\cdot,\cdot,l-1)$, are all identical.
$\Box$\\

First, observe that a periodic FSM with period $l$ can be viewed as
time-invariant FSM in the level of $l$-blocks,
$\{x_{il},x_{il+1},\ldots,x_{il+l-1}\}$, $i=0,1,\ldots$, and hence also in the level of
$\ell$-blocks where $\ell$ is an integer multiple of $l$. Accordingly,
let $\ell$ be an arbitrary integer multiple of $l$, but at the same time,
assume that $\ell$ divides $n$. Denote $m=n/\ell$, and
define the counts,
\begin{equation}
m(z,z',x^\ell)=
\sum_{i=0}^{n/\ell-1}\calI\{z_{i\ell}=z,~z_{i\ell+\ell}=z',~\hat{x}_{i\ell}^{i\ell+\ell-1}=x^\ell\},
~~~~z,z'\in\calS,~x^\ell\in\calX^\ell.
\end{equation}
In fact, there is a certain redundancy in this definition because $z'$ is a
deterministic function of $(z,x^\ell)$ obtained by $\ell$ recursive
applications of the (time-varying) next-state function $g$. 
Concretely, $m(z,z',x^\ell)=m(z,x^\ell)$ iff $z'$ matches
$(z,x^\ell)$ and $m(z,z',x^\ell)=0$ otherwise.
Nonetheless, we
adopt this definition for the sake of clarity of combinatorial derivation to
be carried out shortly. In particular, our derivation will be based on grouping together all
$\{x^\ell\}$, which for a given $z$, yield the same $z'$.
Accordingly, we also denote
$m(z,z')=\sum_{x^\ell\in\calX^\ell}m(z,z',x^\ell)$.

Suppose that the acceptance/rejection criterion that defines $\calA_n$ is based on the counts,
$\{m(z,z',x^\ell),~z,z'\in\calS,~x^\ell\in\calX^\ell\}$, and then the smallest
$\calA_n$ that contains $\bx$ is the type class of $\bx$ pertaining to
$\{m(z,z',x^\ell),~z,z'\in\calS,~x^\ell\in\calX^\ell\}$. The various sequences
in this type class are obtained by permuting distinct $\ell$-tuples
$\{\hat{x}_{i\ell}^{i\ell+\ell-1},~i=0,1,\ldots,m-1\}$ that begin at the same
state, $z$, and end at the same state, $z'$. 
Let us define the empirical distribution,
\begin{equation}
\hat{P}(z,z',x^\ell)=\frac{m(z,z',x^\ell)}{m},~~~~z\in\calS,~a^\ell\in\calX^\ell,
\end{equation}
and the joint entropy,
\begin{equation}
\hat{H}(Z,Z',X^\ell)=-\sum_{z,z',x^\ell}\hat{P}(z,z',x^\ell)\log \hat{P}(z,z',x^\ell),
\end{equation}
and let
\begin{equation}
\hat{H}(X^\ell|Z,Z')=\hat{H}(Z,Z',X^\ell)-\hat{H}(Z,Z'),
\end{equation}
where $\hat{H}(Z,Z')$ is the marginal empirical entropy of $(Z,Z')$.
Then, using the method of types \cite{Csiszar98}, we have:
\begin{eqnarray}
|\calA_n|&\ge&\prod_{z,z'\in\calS}\frac{m(z,z')!}{\prod_{x^\ell\in\calX^\ell}m(z,z',x^\ell)!}\\
&\ge&\prod_{z,z'\in\calS}\left[(m+1)^{-\alpha^\ell}\cdot
\exp_2\left\{\sum_{x^\ell}m(z,z',x^\ell)\log\frac{m(z,z')}{m(z,z',x^\ell)}\right\}\right]\\
&\ge&(m+1)^{-s^2\alpha^\ell}\cdot 2^{m\hH(X^\ell|Z,Z')}\\
&=&(m+1)^{-s^2\alpha^\ell}\cdot 2^{m[\hH(X^\ell)-I(Z,Z';X^\ell)]}\\
&\ge&(m+1)^{-s^2\alpha^\ell}\cdot 2^{m[\hH(X^\ell)-H(Z,Z')]}\\
&\ge&(m+1)^{-s^2\alpha^\ell}\cdot 2^{m[\hH(X^\ell)-2\log s]}\\
&=&\exp_2\left\{n\left[\frac{\hH(X^\ell)}{\ell}-\frac{2\log
s}{\ell}-\frac{\ell s^2\alpha^\ell}{n}\log\left(\frac{n}{\ell}+1\right)\right]\right\},
\end{eqnarray}
and so,
\begin{equation}
R\ge\frac{\log|\calA_n|}{n}\ge\frac{\hH(X^\ell)}{\ell}-\frac{2\log
s}{\ell}-\frac{\ell s^2\alpha^\ell}{n}\log\left(\frac{n}{\ell}+1\right),
\end{equation}
which, for $\ell \gg 2\log s$, can be essentially matched by universal compression for block-memoryless sources and
one-time pad encryption using exactly the same ideas as before. Once again, we have derived a lower bound that is
free of dependence on the
particular FSM, $g$. Recall that $\ell$ divides $n$ and that it is also 
a multiple of $l$, but otherwise, $\ell$ is arbitrary. 
Hence we may maximize this lower bound w.r.t.\ $\ell$ subject to these constraints. 
Alternatively, we may rewrite the lower bound as
\begin{equation}
R\ge\max_{\{q~\mbox{divides}~n/l\}}\left\{\frac{\hH(X^{ql})}{ql}-\frac{2\log
s}{ql}-\frac{ql s^2\alpha^{ql}}{n}\log\left(\frac{n}{ql}+1\right)\right\}.
\end{equation}
Clearly, in view of Remark 5, the above lower bound applies also to the case
of a time-invariant FSM, $g$, but then there would be some mismatch between the upper and lower bound
because to achieve the lower bound, one must gather more detailed empirical statistics, namely, empirical statistics of blocks
together with states, rather than just single symbols with states.

\section{Side Information}
\label{si}

Some of the results presented in the previous sections extend to the case
where side information (SI) is available at the legitimate decrypter and at
the eavesdropper. In principle, it may or may not be available to the
encrypter, and
we consider first the case where it is available.
We assume the SI sequence to be of length $n$, and denote it by
$\by=(y_0,y_1,\ldots,y_{n-1})$, $y_i\in\calY$, $i=0,1,\ldots,n-1$. It is
related to the plaintext, $\bx$, to
be encrypted, but like $\bx$, it is a deterministic, individual sequence.
The SI alphabet $\calY$ is finite and its cardinality, $|\calY|$, is denoted by
$\beta$. Here, the acceptance set depends on $\by$ and denoted accordingly by
$\calA_n(\by)$. The encrypter is therefore a mapping $W=T(\bx,\by,K)$, which
is invertible given $(\by,K)$, allowing the decrypter to reconstruct
$\bx=T^{-1}(W,\by,K)$. 

Consider first a natural extension of Ziv's model for a finite-state
discriminator, which for $i=0,1,\ldots,n-1$, implements the recursion,
\begin{eqnarray}
u_i&=&f(z_i,\hat{x}_i,y_i)\\
z_{i+1}&=&g(z_i,\hat{x}_i,y_i).
\end{eqnarray}
Similarly as before, the acceptance set, $\calA_n(\by)$ is the set of all
$\{\bx\}$, which in the presence of the given $\by$, yield the all-zero
output sequence, $\bu=(u_0,u_1,\ldots,u_{n-1})=(0,0,\ldots,0)$.

Let $c(\bx,\by)$ denote the number of phrases of joint parsing of
$$(\bx,\by)=((x_1,y_1),(x_2,y_2),\ldots,(x_n,y_n))$$ 
into distinct phrases,
and let $c_l(\bx|\by)$
denote the number of occurrences of $\by(l)$ - the $l$-th distinct phrase of $\by$, which
is also the number of different phrases of $\bx$ that appear jointly with
$\by(l)$. Let $c(\by)$ denote the number of different $\{\by(l)\}$, that is,
$\sum_{l=1}^{c(\by)}c_l(\bx|\by)=c(\bx,\by)$. 
Finally, let $c_{lzz'}(\bx|\by)$ denote the number of $\bx$-phrases
that are aligned with $\by(l)$, beginning at state $z$ and ending at state $z'$.
Then, similarly as in the derivation in \cite{Ziv78u} (see also Appendix), 
\begin{equation}
|\calA_n(\by)|\ge \prod_{l=1}^{c(\by)}\prod_{z,z'}c_{lzz'}(\bx|\by)^{c_{lzz'}(\bx|\by)},
\end{equation}
and so, defining the auxiliary RV's $(Z_l,Z_l')$, $l=1,\ldots,c(\by)$, as being jointly distributed
according to 
\begin{equation}
Q_l(z,z')=\frac{c_{lzz'}(\bx|\by)}{c_l(\bx|\by)}, ~~~z,z\in\calS,
\end{equation}
we have
\begin{eqnarray}
\log|\calA_n(\by)|&\ge&\sum_{l=1}^{c(\by)}\sum_{z,z'}c_{lzz'}(\bx|\by)\log
c_{lzz'}(\bx|\by)\nonumber\\
&=&\sum_{l=1}^{c(\by)}c_l(\bx|\by)\sum_{z,z'}\frac{c_{lzz'}(\bx|\by)}{c_l(\bx|\by)}\left[\log
\frac{c_{lzz'}(\bx|\by)}{c_l(\bx|\by)}+\log c_l(\bx|\by)\right]\nonumber\\
&=&\sum_{l=1}^{c(\by)}c_l(\bx|\by)\log c_l(\bx|\by)
-\sum_{l=1}^{c(y)}c_l(\bx,\by)H(Z_l,Z_l')\nonumber\\
&\ge&\sum_{l=1}^{c(\by)}c_l(\bx|\by)\log c_l(\bx|\by)
-2\cdot\sum_{l=1}^{c(y)}c_l(\bx,\by)\log s\nonumber\\
&=&\sum_{l=1}^{c(\by)}c_l(\bx|\by)\log c_l(\bx|\by)
-2c(\bx,\by)\log s\nonumber\\
&\ge&\sum_{l=1}^{c(\by)}c_l(\bx|\by)\log c_l(\bx|\by)
-\frac{2n\log s}{(1-\epsilon_n)\log n},
\end{eqnarray}
where the last inequality follows from \cite{ZL78} (see also \cite[Lemma
13.5.3]{CT06}) with $\epsilon_n=O\left(\frac{\log(\log n)}{\log n}\right)$.
This lower bound can be asymptotically attained by the conditional version of
the LZ algorithm (see \cite{Ziv85} and \cite{UK03}),
followed by one-time pad encryption.
If $\by$ is unavailable at the encrypter, $\bx$ can still be compressed into
about 
\begin{equation}
u(\bx|\by)=\sum_{l=1}^{c(\by)}c_l(\bx|\by)\log c_l(\bx|\by) 
\end{equation}
bits before the one-time pad
encryption, using
Slepian-Wolf encoding \cite[Section 15.4]{CT06} and reconstructing (with high
probability) after decryption using a universal
decoder that uses $u(\bx|\by)$ as a decoding metric, see \cite{me24}.

Unfortunately, and somewhat surprisingly, a direct extension of the results of
Subsections \ref{fsm+c} and \ref{srm+c} to the case with SI turns out to be
rather elusive.
The reason is the lack of a single-letter formula for the exponential growth
rate of the cardinality of a finite-state conditional type class of $\bx$-vectors that is defined by joint counts
of the form $n(x,y,z)=\sum_{i=1}^n\calI\{x_i=x,y_i=y,z_i=z\}$, even in the
simplest special case where $z_i=x_{i-1}$. In a nutshell,
this quantity depends on $\by$ in a complicated manner, which cannot be
represented in terms of an empirical distribution of fixed dimension that
does not grow with $n$. However, we can obtain at least a lower bound if we
treat these cases as special cases of the periodically time-varying FSMs with
counters in view of Remark 5.

Finally, consider the class of discriminators implementable by
periodically time-varying FSMs with SI, defined as follows:
\begin{eqnarray}
z_{i+1}&=&g(z_i,\hat{x}_i,y_i,i~\mbox{mod}~l),
\end{eqnarray}
where $i=0,1,2,\ldots$ and $l$ is as in Subsection \ref{ptv+c}.
The extension of the derivation in Subsection \ref{ptv+c} is quite
straightforward - one has to count permutations of $\ell$-vectors of $\bx$
that, not only share the same initial and final states, but are also aligned to the same
$\ell$-blocks of $\by$. The resulting bound is given by:
\begin{equation}
R\ge\frac{\log|\calA_n(\by)|}{n}\ge\frac{\hat{H}(X^\ell|Y^\ell)}{\ell}-\frac{2\log
s}{\ell}-\frac{\ell s^2\alpha^\ell\beta^\ell}{n}\log\left(\frac{n}{\ell}+1\right),
\end{equation}
which, for $\ell \gg 2\log s$, can be essentially matched by universal
compression for block-memoryless sources in the presence of SI and
one-time pad encryption. In the absence of SI at the encrypter, one may use
universal Slepian-Wolf coding for block-memoryless sources, which is a direct
extension of universal Slepian-Wolf coding for memoryless sources, see, e.g.,
\cite{Draper04}.

\section*{Appendix - Proof of the Last Step of Eq.\ (\ref{R>LZ})}
\renewcommand{\theequation}{A.\arabic{equation}}
    \setcounter{equation}{0}

Since reference \cite{Ziv78u} is unpublished, then for the sake of
completeness, we provide here the essential steps of Ziv's proof of the inequality,
\begin{equation}
\log|\calA_n|\ge LZ(\bx)-n\epsilon_n,
\end{equation}
which must hold for every finite-state discriminator with no more than $s$
states, where $\epsilon_n= O\left(\frac{\log(\log n)}{\log n}\right)$ for
fixed $s$. The final steps are different from
those of \cite{Ziv78u} by adopting a somewhat simpler approach, that is presented in
Section 13.5 of \cite{CT06}.

Let $\bx$ be parsed into $c$ distinct phrases, 
$$(x_0,x_1,\ldots,x_{n_1-1}),
(x_{n_1},x_{n_1+1},\ldots,x_{n_2-1}),\ldots,
(x_{n_c},x_{n_c+1},\ldots,x_{n-1})$$
with the possible exception of the last phrase which might be incomplete,
and let $c_{lzz'}$ denote the number of
phrases of length $l$ wherein the initial state of the FSM  $g$
is $z$ and the final state is $z'$ (of course, $\sum_{l,z,z'}c_{lzz'}=c$). If
the discriminator
accepts $\bx$, namely, if $\bu=(0,0,\ldots,0)$, then for every $(l,z,z')$, each
of the $c_{lzz'}$ phrases can
be replaced by any of the other such phrases to
obtain new sequences of length $n$ for which the output is also $\bu=(0,0,\ldots,0)$, and
hence must be accepted too.
Now, let $(L,Z,Z')$ denote a triple of
auxiliary random variables jointly distributed according to
the probability distribution 
\begin{equation}
Q(l,z,z')=\frac{c_{lzz'}}{c},~~~~~~z,z'\in\calS,~l=1,2,\ldots
\end{equation}
and let $H(L,Z,Z')$ denote the joint entropy of $(L,Z,Z')$.
Then,
\begin{equation}
|\calA_n|\ge \prod_{l,z,z'}(c_{lzz'})^{c_{lzz'}},
\end{equation}
where $(c_{lzz'})^{c_{lzz'}}$ on the right-hand side is the number of ways each one of the $c_{lzz'}$
phrases of length $l$, starting at state $z$ and ending at state $z'$ can be
replaced by any other phrase with the same qualifiers. 
Consequently,
\begin{eqnarray}
\log|\calA_n|&\ge&\sum_{l,z,z'}c_{lzz'}\log c_{lzz'}\\
&=&c\cdot\sum_{l,z,z'}\frac{c_{lzz'}}{c}\left[\log \frac{c_{lzz'}}{c}+\log
c\right]\\
&=&c\log c-c\cdot H(L,Z,Z')\\
&\ge&c\log c-c[H(L)+H(Z)+H(Z')]\\
&\ge&c\log c-c\cdot H(L)-2c\log s.
\end{eqnarray}
Now, the entropy of $L$, given that $\bE\{L\}=n/c$, cannot be larger than
$(n/c+1)\log(n/c+1)-(n/c)\log(n/c)\le 1+\log(n/c+1)$ (see, Lemma 13.5.4 and
eqs.\ (13.120)-(13.122) in \cite{CT06}). Thus,
\begin{equation}
\log|\calA_n|\ge c\log c-2c\log s-c-c\log\left(\frac{n}{c}+1\right),
\end{equation}
and then
\begin{eqnarray}
\label{last}
R&\ge&\frac{\log|\calA_n|}{n}\nonumber\\
&\ge&\frac{c\log c}{n}-\frac{2c}{n}\log
s-\frac{c}{n}\log\left(\frac{n}{c}+1\right)\nonumber\\
&=&\frac{c\log c}{n}-\frac{2c}{n}\log
s-O\left(\frac{\log\log n}{\log n}\right),
\end{eqnarray}
where the last line is obtained similarly as in eqs.\ (13.123)-(13.124) in
\cite{CT06}. The final step of eq.\ (\ref{R>LZ}) is obtained from the fact
that $LZ(\bx)$ is upper bounded by $c\log c$ plus terms that, after
normalizing by $n$, are
are negligible compared to $\frac{\log(\log n)}{\log n}$ -- see Theorem 2 in \cite{ZL78}.

\end{document}